\documentstyle[aps,eqsecnum,preprint]{revtex}

\begin{document}

\draft

\title{Wave functions for arbitrarily Polarized
	Quantum Hall States}
\author{Sudhansu S. Mandal \cite{e1} and V. Ravishankar
\cite{e2}}
\address{Department of Physics, Indian Institute of technology,
Kanpur -- 208 016, INDIA }

\maketitle

\begin{abstract}

We determine the wave functions for
arbitrarily polarized quantum Hall states by employing
the doublet model which has been
proposed recently to describe arbitrarily polarized quantum Hall
states. Our findings recover the well known fully
polarized Laughlin wave functions and unpolarized Halperin wave
function for the filling fraction $\nu =2/5$.
We have also confirmed by an
explicit One-loop computation that the Hall conductivity does
indeed get quantized at those filling fractions that follow from
the model. Finally, we have given a physical picture for the
non-analytic nature of the wave functions, and shown that quantum
fluctuations restore the Kohn mode.

\end{abstract}



\section{INTRODUCTION}

Recently Fradkin \cite{frad} has proposed a method for determining
the absolute square of the many particle wave function for the
ground state of a system
from a knowledge of correlation functions which are
generally computed in field theory. This beautiful method which
employs the generating functions for equal-time correlation functions
is applicable to any field theoretic problem. If one further knows
before hand, or assumes, that the many
particle system is non-degenerate,
the many particle wave function is also determined thereby, apart from
the gauge dependent phase which is essential in expectation values
of observables such as velocity. Using this formalism, Lopez and
Fradkin \cite{lopez1} have extracted the wave functions for
fully polarized quantum Hall (QH) states
within the composite fermion model (CFM)
\cite{jain1,jain2}.
Remarkably, they recover the Jastrow part of the Laughlin
wave functions \cite{laugh} unambiguously
for filling fractions $\nu =1/(2k +1)$
($k$ an integer) which have been shown to be exact by
Kivelson and Trugman \cite{trug} and Haldane \cite{hald}.
Note that this result vindicates the mean field (MF) ansatz which
one normally employs for the Chern-Simons (CS) field. More
generally, Lopez and Fradkin \cite{lopez1} show that the Jastrow
form is indeed generic to all states with filling fractions
$\nu = p/(2sp \pm 1)$ ($p,\,s$ are integer) which occur in CFM.
Further, the long distance properties of the many body wave
function are universal, being independent of the microscopic
charge-charge interaction. Yet another interesting aspect is
that the wave functions are in general non-analytic. That is,
the exponents that occur in the Jastrow form are arbitrary
rational numbers. (We discuss the origin of the non-analyticity
in section--III-C).

The above analysis of Lopez and Fradkin \cite{lopez1} is based
on the strict assumption that the
spin degree of freedom of the electrons
is frozen in the direction of the magnetic field, i.e., they are
assumed to be spinless. Such an
assumption which is valid for large magnetic fields $(B\sim 10
T)$ breaks down for small values of $B$ :
the Zeeman splitting is
now not that large (also partly due to small $g$ value, $g\sim 0.4$)
and, as Halperin \cite{halp} has observed, these systems are not fully
polarized. Indeed, experiments reveal that at relatively
small values of $B$, the QH states at filling factors $\nu = {4 \over
3}, \, {8 \over 5},\, {10 \over 7}$
(Ref. \cite{clark,eisen1}) and ${2 \over 3}$
(Ref. \cite{eisen2,engel}) are unpolarized
while the states at $\nu = {3 \over 5}$ (Ref. \cite{engel})
and ${7 \over 5}$
(Ref. \cite{clark}) are partially
polarized. Further, it is also known experimentally
that the states
which are at maximum polarization to start
with pass over to partially polarized
or unpolarized states as the Zeeman energy is lowered sufficiently ---
either by reducing the tilting angle of the magnetic field
\cite{clark,eisen1,eisen2,engel} or
by decreasing the electron density \cite{eisen2}.
In the vanishing Zeeman splitting (VZS)
limit, it has been found  from numerical computations \cite{tchak1}
that the states with
$\nu =2/(2n+1)$ are unpolarized and those of the Laughlin sequence
\cite{laugh} with $\nu =1/(2n+1)$ are fully polarized, in the
thermodynamic limit. Also the state at $\nu ={3 \over 5}$ has been
found to be partially polarized by an exact
diagonalization study \cite{tchak2},
in agreement with experiments.

Wu, Dev and Jain \cite{wu} have studied this problem and constructed
trial wave functions by employing the CFM.
These trial wave functions
are confirmed to be exact by numerical computation.
They report that, in the VZS limit, all even numerator QH states
are unpolarized and all those states with both the numerator and
denominator (of $\nu$) odd are partially/fully polarized. Further,
Belkhir and Jain \cite{belk} have proposed
that the CFM accommodates
the sequence $\nu =2n/(3n+2)$ all of which are spin unpolarized.
 From the wave functions that they construct, they also interpret that
these states possess a new feature where each composite fermion
carries two different types of vortices --- one of which seen by
all electrons while the other is visible only to an electron
of like spin.

Recently we have proposed \cite{spin} a global model which
employs a doublet \cite{wen} of CS gauge fields, with the strength
of the CS term given by a symmetric coupling matrix.
The model is within the composite fermion framework, i.e., each
fermion has an even number of vortices attached to it.
The model accounts
for all the observed as well as proposed \cite{wu} QH states
with the correct spin polarization. Further it predicts new
possible QH states characterized by two gaps (corresponding to
two spin species) of excitations.

In this context we mention that a very similar model has been
proposed by Lopez and Fradkin \cite{lopez4} to describe QH
effects in double layered systems. In comparing the two models,
we note that while the same kind of the CS action
is employed in both the models, the problems addressed are otherwise
of entirely different nature. The problem at hand is the
description of QH systems with the spin degree of freedom,
restricted to a single layer; Lopez and Fradkin \cite{lopez4}
study spinless fermions in double layered systems.
This leads to different consequences. For instance, in the
model of Lopez and Fradkin,
``the spin singlet state
(3,3,2), which has filling fraction $\nu =2/5$, cannot be
described within the Abelian Chern-Simons approach"\cite{lopez4}.
On the other
hand,
the $2/5$ state which is the most important for us here emerges
naturally in our model. More importantly, the inclusion of the
spin degree of freedom in the paper of Lopez and Fradkin
\cite{lopez4} involves the more complicated non-abelian SU(2) CS
model, which was first introduced by Frohlich et al \cite{frohl}
in CS action to study QH systems.
The inclusion of spin degree is only to the extent of describing
the singlet states, unlike the present paper where arbitrary
polarization are discussed. In that sense, we claim that the
present model provides a more comprehensive and a simpler
picture than the non-abelian models employed in Ref.~\cite{lopez4}.
A detailed comperative study between our approach and results
with those of Lopez and Fradkin \cite{lopez4} will be provided in
the Appendix.

It is the purpose of this paper to extract wave functions for
arbitrary polarized states by employing the doublet model
\cite{spin}. For simplicity's sake, we shall restrict to only
those sequences which have been seen significantly. They
correspond to states with a single gap of excitation.
We show that the asymptotic properties of the wave
function, in the VZS as well as
thermodynamic limits, are completely determined by the long-distance
behaviour of the equal-time density-density correlation functions of
different spin species of electrons. We explicitly determine the
square of the absolute value of the wave function for
(i) spin-unpolarized,
(ii) partially polarized and (iii) fully polarized QH states.
Gratifyingly, we recover the well-known
wave function for spin-unpolarized state $\nu = {2 \over 5}$
which was predicted by Halperin \cite{halp}, in our
analysis.
All the unpolarized states, having
numerator 2, emerge as analytic functions (which is not the case for
spinless electrons). However, the wave functions for integer QH states
with $\nu >2$ and all the fractional QH states having numerator greater
than 2 continue to remain non-analytic. Interestingly we also find
that, for the integer QH states, the particles
with dissimilar spins are completely uncorrelated.
Finally, we predict that the wave function for even denominator
unpolarized QH states are analytic in nature.

In the next section we briefly discuss the doublet model and
determine the equal-time generating functional for mixed spin
density-density correlations. Section 3 is devoted to a
determination of the wave functions and the origin of
non-analyticity is briefly discussed. Finally in section 4, we
perform the linear response analysis to demonstrate explicitly
the quantization of Hall conductivity $\sigma_H$ at appropriate
filling fractions and show that the Kohn-mode is restored by the
fluctuations. We conclude the paper in section 5.

\section{The Generating Functional}

\subsection{The Doublet Model}

We first briefly discuss the model \cite{spin} employed here.

Consider a two-dimensional system of non-relativistic
spin-1/2 interacting fermions in the presence of magnetic field
perpendicular to the plane. In their study of spinless fermions,
Lopez and Fradkin \cite{lopez2} have shown that such a system is
equivalent to the one interacting with a CS gauge field provided
the CS parameter is such that the statistics of the particles
remains fermionic. Using
this generic argument, we propose a generalized Lagrangian density
\begin{eqnarray}
{\cal L} &=& \psi_\uparrow^\dagger {\cal D}
     (A_\mu^\uparrow +a_\mu^\uparrow )\psi_\uparrow
   + \psi_\downarrow^\dagger {\cal D}(A_\mu^\downarrow +
   a_\mu^\downarrow )\psi_\downarrow
   +{1 \over 2} \tilde{a}_\mu\epsilon^{\mu\nu \lambda } \Theta
   \partial_\nu a_\lambda \nonumber \\
   & & -eA_0^{\rm in}\rho +{1 \over 2}\int d^3 x^\prime A_0^{\rm
   in}(x)V^{-1}(x-x^\prime)A_0^{\rm in} (x^\prime)\; .
\label{eq1}
\end{eqnarray}
Here $\psi $ is the fermionic field and $\uparrow (\downarrow )$
represents spin-up (down),
\begin{equation}
{\cal D}(A_\mu^r +a_\mu^r) = iD_0^r +(1/2m^\ast)
D_k^{r\, 2} +\mu +(g/2)\mu_B (B+B^r+b^r)\sigma \; ,
\label{eq2}
\end{equation}
with $D_\mu^r =\partial_\mu -ie
(A_\mu +A_\mu^r+a_\mu^r)$ where $A_\mu $
is the external electro-magnetic field which interacts with all
the electrons while $A_\mu^r$ and $a_\mu^r$ are the external
probe \cite{fnote1} and the CS gauge field
respectively, interacting with {\it only} the particles having
spin indices $r= \uparrow \, ,\, \downarrow $. The field
$A_0^{\rm in}$
is identified as an internal scalar potential. Particles
with an effective mass $m^\ast$ and charge $e$ have mean density
$\rho$ which is fixed by the introduction of chemical potential
$\mu$ as a Lagrange multiplier. (We have chosen the units $\hbar
=c =1)$. Note that the Zeeman term includes all the three kinds
of magnetic fields, $\mu_B$ is the Bohr-magneton, and $\sigma =+1
(-1)$ for spin-up (down) electrons. We have introduced
a doublet of CS gauge fields in (\ref{eq1}) as
\begin{equation}
a_\mu = \left( \begin{array}{c}
    a_\mu^\uparrow \\
    a_\mu^\downarrow \end{array}
    \right) \; ,
\label{eq3}
\end{equation}
and the strength of the CS parameter is taken to be
\begin{equation}
\Theta = \left( \begin{array}{cc}
    \theta_1 & \theta_2 \\
    \theta_2 & \theta_1 \end{array}
    \right)  \; .
\label{eq4}
\end{equation}
$\tilde{a}_\mu $ is the transpose of the doublet field $a_\mu$.
The fourth term in Eq.~(\ref{eq1}) describes the charge
neutrality of the system. Finally, $V^{-1}(x-x^\prime )$
is the inverse
of the electron interaction potential (in the operator sense).
The usual fermion interaction term in quartic form would be
achieved by an integration over $A_0^{\rm in}$ field. The
action given by Eq.~(\ref{eq1}) is invariant under the gauge
transformations $a_\mu^{ \uparrow , \downarrow } \rightarrow
a_\mu^{ \uparrow , \downarrow } + \partial_\mu \lambda ^{
\uparrow , \downarrow } (x)\, ,\, \psi_{ \uparrow , \downarrow }
(x) \rightarrow  \exp [ie \lambda^{ \uparrow  , \downarrow }
(x)] \psi_{ \uparrow  , \downarrow } (x)$.
In other words, the doublet model is abelian.

We diagonalize the matrix $\Theta $, with the eigen values
$ \theta_\pm = \theta_1 \pm \theta_2$, and denote $a_\mu$ in the
eigen basis by
\begin{equation}
a_\mu =  \left(
\begin{array}{c}
a_\mu^+ \\
a_\mu^-
\end{array}
\right)  \; .
\end{equation}
By simple rescalings, Eq.~(\ref{eq1}) may be written as
\begin{eqnarray}
{\cal L} &=& \psi_\uparrow^\dagger {\cal D}
(A_\mu^\uparrow +a_\mu^+ +a_\mu^-)\psi_\uparrow
  + \psi_\downarrow^\dagger {\cal D} (A_\mu^\downarrow +
  a_\mu^+ -a_\mu^-)\psi_\downarrow
  +{\theta_+ \over 2 }\epsilon^{\mu\nu \lambda}
  a_\mu^+\partial_\nu a_\lambda^+ \nonumber \\
 & & +{\theta_- \over 2 }\epsilon^{\mu\nu \lambda}
  a_\mu^-\partial_\nu a_\lambda^-
    -eA_0^{\rm in}\rho +{1 \over 2}\int d^3 x^\prime A_0^{\rm
   in}(x)V^{-1}(x-x^\prime)A_0^{\rm in} (x^\prime)\; .
 \label{eq6}
\end{eqnarray}
This incorporates the idea that each electron, in general, has
two kinds of vortices associated with it --- while the
contributions of vortices are added for spin up particles, the
spin down particles get their subtracted contribution.

\subsection{Case--1: No Polarization}

We first study the case $ \theta_1 = \theta_2 = \theta $ (say).
Here $ \theta_+ =2 \theta $ and $ \theta_- =0$. Hence the gauge
field $a_\mu^-$ decouples dynamically and merely plays the role
of a Lagrange multiplier: $( \partial {\cal L} / \partial
a_0^- )= \rho_ \uparrow - \rho_ \downarrow \equiv 0$, where
$\rho_ \uparrow (\rho_ \downarrow )$ is the density for spin-up
(down) particles. Thus the unpolarized case is accomplished by
the choice $ \theta_1 = \theta_2$. Rescaling $ \theta $ by $
\theta /2 $, we parametrize $ \theta = (e^2 /2\pi)(1/2s)$ ($s$
is an integer) in order to impose the composite fermion picture.

The generating functional of the system (in uniform magnetic
field $B$) can be written as
\begin{equation}
{\cal Z} \left[ A_\mu^\uparrow , A_\mu^\downarrow  \right] =
\int [d\psi^\dagger_ \uparrow ][d\psi_ \downarrow
][d\psi^\dagger_ \downarrow ][d\psi_ \downarrow ][da_\mu^+]
[dA_0^{\rm in}] e^{i\int d^3 x {\cal L}} \; .
\label{eq7}
\end{equation}
The terms (quadratic) in $\psi_\uparrow $ and $\psi_\downarrow $
fields can be integrated out to produce the fermion determinants
for spin-up and down respectively. The other terms remain as
they are in the Lagrangian density (\ref{eq6}). Now the fermion
determinants are to be expanded about a saddle point of gauge
fields which we fix as follows.

It is clear that the
fermions are associated with an even number $(2s)$ of flux
quanta. In the mean field (MF) ansatz, these fluxes produce an
average CS magnetic field $ \langle b^+ \rangle = -e\rho /
\theta $ which is seen by all the electrons. Demanding that the
effective Landau levels (LL) formed by the effective magnetic
field $\bar{B}^+ =B+ \langle b^+ \rangle $ accommodate all the
particles at an even integer filling factor $2p$, ($p$ for up
spin and $p$ for down), the actual filling fraction $\nu$ is
obtained as
\begin{equation}
\nu = {2p \over 4sp +1} \; .
\label{eq8}
\end{equation}
The energy corresponding to each level is obtained as
$\varepsilon_{n\sigma} =(n+1/2)\bar{\omega}_c -{g \over 2} \mu_B
\bar{B}^+ \sigma$ ($n=0,1,\ldots $),
where the effective cyclotron frequency
$\bar{\omega}_c ={e \over m^\ast}\bar{B}^+$.
The actual cyclotron frequency $\omega_c$ of the system is
related to $\bar{\omega}_c$ via $\omega_c =\bar{\omega}_c
(1+4sp)$. Recall that $p$ can be a negative integer (meaning
$\bar{B}^+$ is antiparallel to $B$). All the states obeying
Eq.~(\ref{eq8}) are spin unpolarized. It is known \cite{lopez2}
that a liquid-like solution exists for the vanishing average of
$A_0^{\rm in}$ field. Therefore the saddle point is fixed at
$\langle b^+ \rangle =-e\rho / \theta $ and $\langle A_0^{\rm
in} \rangle =0$.
We remark that this model does not accommodate the sequence $\nu
=2n/(3n+2)$ proposed by Belkhir and Jain \cite{belk}. Of course,
our model does include all the unpolarized states that are seen
experimentally.

Expanding the fermion determinants about the above mentioned
saddle point, up to terms quadratic in the gauge field
fluctuations and the external probes, we obtain
\begin{equation}
{\cal Z}\left[ A_\mu^\uparrow , A_\mu^\downarrow \right] = \int
[da_\mu^+][dA_0^{\rm in}] \exp \left[ iS_{\rm eff} \left(
A_\mu^\uparrow ,A_\mu^\downarrow ,a_\mu^+, A_0^{\rm in} \right)
\right] \; ,
\label{eq9}
\end{equation}
where $S_{\rm eff}$ is identified as one-loop effective action
and has the form
\begin{eqnarray}
S_{\rm eff} &=&
   -{1 \over 2}\int d^3x \int d^3 x^\prime (A_\mu^\uparrow
   +a_\mu^+ +A_0^{\rm in}\delta_{\mu 0})(x)\Pi^{\mu\nu}_
   \uparrow (x,x^\prime)(A_\nu^\uparrow
   +a_\nu^+ +A_0^{\rm in}\delta_{\nu 0})(x^\prime) \nonumber \\
 & &   -{1 \over 2}\int d^3x \int d^3 x^\prime (A_\mu^\downarrow
   +a_\mu^+ +A_0^{\rm in}\delta_{\mu 0})(x)\Pi^{\mu\nu}_
   \downarrow (x,x^\prime)(A_\nu^\downarrow
   +a_\nu^+ +A_0^{\rm in}\delta_{\nu 0})(x^\prime) \nonumber \\
 & & +\int d^3x {\theta \over 2 }\epsilon^{\mu\nu \lambda}
  a_\mu^+\partial_\nu a_\lambda^+
    +{1 \over 2}\int d^3x \int d^3 x^\prime A_0^{\rm
   in}(x)V^{-1}(x-x^\prime)A_0^{\rm in} (x^\prime)\; .
 \label{eq10}
\end{eqnarray}
$a_\mu^+$ and $A_0^{\rm in}$ now represent fluctuations about
their corresponding mean values. The polarization tensors
$\Pi^{\mu \nu}_{ \uparrow , \downarrow }$ will be evaluated
below.

The procedure for the evaluation of polarization tensor is well known
in literature (see e.g., \cite{lopez2}) for spinless particles.
The additional feature is that, here we evaluate
$\Pi^{\mu \nu}_{ \uparrow , \downarrow }$ for two different spin
species respectively.
In brief, it follows from translational and gauge
invariance that $\Pi^{\mu \nu}_{ \uparrow , \downarrow }$
have the form (in momentum space)
\begin{eqnarray}
\Pi^{\mu \nu}_{ \uparrow , \downarrow } &=&
\Pi_0^{\uparrow , \downarrow } (\omega\, ,\, {\bf q}^2)
(q^2g^{\mu\nu} -q^\mu q^\nu )+ \left( \Pi_2^{\uparrow ,
\downarrow}-\Pi_0^{\uparrow , \downarrow} \right) (\omega \, ,
\, {\bf q}^2)  \nonumber \\
& & \times \left( {\bf q}^2 \delta^{ij} -q^{i}q^{j} \right)
\delta^{\mu i}\delta^{\nu j} +i\Pi_1^{\uparrow , \downarrow}
(\omega \, ,\, {\bf q}^2) \epsilon^{\mu\nu \lambda} q_ \lambda
\; . \label{eq11}
\end{eqnarray}
The form factors are then evaluated in the lowest order in ${\bf
q}^2$ and we find
\begin{mathletters}
\label{eq12}
\begin{equation}
\Pi_0^{ \uparrow , \downarrow}= -{e^2 p \over 2\pi} {\bar{\omega}_c
\over \omega^2 -\bar{\omega}_c^2} \equiv \Pi_0 \; \; ; \; \;
\Pi_1^{ \uparrow , \downarrow}= \Pi_0^{\uparrow , \downarrow}
\bar{\omega}_c \equiv \Pi_1 \; ;
\end{equation}
\begin{equation}
\Pi_2^\uparrow =  -{e^2 \over 4\pi m^\ast} \bar{\omega}_c^2
\left[ {3 \over \omega^2 -\bar{\omega}_c^2} -
{4 \over \omega^2 -4\bar{\omega}_c^2 } \right] p(p-1)\; ;
\end{equation}
\begin{equation}
\Pi_2^\downarrow =  -{e^2 \over 4\pi m^\ast} \bar{\omega}_c^2
\left[ {3 \over \omega^2 -\bar{\omega}_c^2} -
{4 \over \omega^2 -4\bar{\omega}_c^2 } \right] p(p+1)\; .
\end{equation}
\end{mathletters}

To evaluate the effective action for the external probes
$A_\mu^{\uparrow , \downarrow}$, we integrate over the internal
fields $a_\mu^+$ and $A_0^{\rm in}$ in (\ref{eq9}). Thus we
obtain (in momentum space),
\begin{equation}
S_{eff} \left[ A_\mu^\uparrow ,A_\mu^\downarrow \right] = {1
\over 2} \int {d^3 q \over (2\pi)^3} A_\mu^r (q)
K^{\mu\nu}_{rr^\prime} (\omega \, ,\, {\bf q}^2)
A_\nu^{r^\prime} (-q) \; ,
\label{eq13}
\end{equation}
where the indices $r,r^\prime = \uparrow , \downarrow $.
Since we need only the density-density correlations between
different spin species, it is sufficient that we evaluate
$K^{00}_{rr^\prime}$. We find (for small ${\bf q}^2$)
\begin{mathletters}
\label{eq14}
\begin{eqnarray}
K^{00}_{\uparrow \uparrow}=K^{00}_{\downarrow \downarrow} & = &
{1 \over 2} \left[ \Pi_0 -{\Pi_0 \theta^2 \over 4
\Pi_0^2\omega^2 -(2\Pi_1 + \theta)^2 }\right] {\bf q}^2 +{\cal
O} (({\bf q}^2)^2) \; ; \\
K^{00}_{\uparrow \downarrow}=K^{00}_{\downarrow \uparrow} & = &
-{1 \over 2} \left[ \Pi_0 +{\Pi_0 \theta^2 \over 4
\Pi_0^2\omega^2 -(2\Pi_1 + \theta)^2 }\right] {\bf q}^2 +{\cal
O} (({\bf q}^2)^2) \; ,
\end{eqnarray}
\end{mathletters}
subject to the condition $\lim_{{\bf q}^2 \rightarrow  0} V({\bf
q }^2) {\bf q}^2 =0$. In other words, Eq.~(\ref{eq14}) is valid
for any potential which is short ranged compared to $\ln r$.
Here $K^{00}_{\uparrow
\uparrow}$, $K^{00}_{\uparrow \downarrow}$, $K^{00}_{\downarrow
\uparrow}$ and $K^{00}_{\downarrow \downarrow}$ represent the
density-density correlations among spin up-up, up-down, down-up
and down-down species of the particles respectively. We
next write the generating functional for equal-time
density-density correlations among mixed spins in the form,
\begin{equation}
{\cal Z} \left[ A_0^\uparrow ,A_0^\downarrow\right] =\exp\left[
-{1
\over 2}\int {d^2 {\bf q} \over (2\pi)^2}A_0^r ({\bf q})
{\cal K}_{rr^\prime}A_0^{r^\prime}(-{\bf q}) \right]
\label{eq15}
\end{equation}
with
\begin{equation}
{\cal K}_{rr^\prime}=\int {d\omega \over 2\pi i} K^{00}_{rr^\prime}
(\omega\, ,\, {\bf q}^2)
\label{eq16}
\end{equation}
which is given by
\begin{equation}
{\cal K}= {{\bf q}^2 \over 2\pi}({e^2 \over 2}){p \over 4sp+1} \left[
\begin{array}{cc}
  2sp+1 & -2sp \\
  -2sp & 2sp+1
 \end{array}
 \right]
 \; . \label{eq17}
 \end{equation}

\subsection{Case--2:  Partial Polarization}

Now consider the case $\theta_1 \neq \theta_2$, i.e., $\theta_\pm
\neq 0$. The generating functional, in this case, reads
\begin{equation}
{\cal Z} \left[ A_\mu^\uparrow , A_\mu^\downarrow  \right] =
\int [d\psi^\dagger_ \uparrow ][d\psi_ \downarrow
][d\psi^\dagger_ \downarrow ][d\psi_ \downarrow ][da_\mu^+]
[da_\mu^-][dA_0^{\rm in}] e^{i\int d^3 x {\cal L}} \; ,
\label{eq18}
\end{equation}
where there is an additional functional integral over the field
$a_\mu^-$.
The fermionic fields are then integrated out to produce the
fermion determinants which are to be expanded about the saddle
point of the gauge fields. We parametrize $\theta_\pm =(e^2
/2\pi)(1/s_\pm)$. As discussed in Ref. \cite{spin},
we choose $s_-=0$. i.e., mean CS magnetic
field $\langle b^- \rangle =0$.
In this case, the field $a_\mu^-$ provides a vanishing mean
field $\langle b^- \rangle$, and does not contribute at tree
level (in contrast to the unpolarized case where $a_\mu^-$
is completely non-dynamical). We impose the composite fermion
requirement, which fixes $s_+ =2s$ (an even integer) \cite{spin}. Since
$\langle b^- \rangle =0$, the effective magnetic field for both
spin up and down particles is same and is given by $\bar{B}^+
=B+\langle b^+ \rangle $ where $\langle b^+ \rangle =-e\rho
/\theta_+$ in the MF ansatz. Let $p_ \uparrow (p_ \downarrow )$
be the number of effective LL formed by $\bar{B}^+$ filled by up
(down) particles. This leads to the actual filling fraction
and the spin density to be
\begin{equation}
\nu = {p_ \uparrow + p_ \downarrow \over 2s(p_ \uparrow + p_
\downarrow ) +1}\;\; ; \; \;
\Delta\rho = \rho \left( {p_ \uparrow -p_ \downarrow
\over  p_ \uparrow + p_ \downarrow } \right) \; .
\label{eq19}
\end{equation}
Note that $p_ \uparrow $ and
$p_ \downarrow $ can be negative integers as well (when
$\bar{B}^+$ is antiparallel to $B$).
Further, it is easy to check that the choice $\Delta\rho =0$,
i.e., $p_ \uparrow = p_ \downarrow $ simply collapses to case-1.
We thus require that $p_ \uparrow
\neq p_ \downarrow $, which naturally leads to partial
polarization. The effective cyclotron frequency $\bar{\omega}_c$
is related to $\omega_c = eB/m^\ast $ by
$\omega_c =\bar{\omega}_c [2s(p_\uparrow +p_
\downarrow ) +1 ]$.
For small Zeeman energies, we may take $p_ \uparrow = p_
\downarrow +1 = p$ (say). Then
\begin{equation}
{\Delta\rho \over \rho} = {1 \over 2p-1} \;\; ;\;\;
\nu = {2p-1 \over 2s(2p-1)+1} \; .
\label{eq20}
\end{equation}
The sequence (\ref{eq20}) is indeed partially polarized
becoming fully polarized for $p=1$.
Then $\nu_{p=1}=1/(2s+1)$ which is simply the
Laughlin sequence \cite{laugh} known theoretically
to be fully polarized
\cite{tchak1}. In short, for obtaining partially/fully polarized
QH states we fix the saddle point at $\langle b^+ \rangle =
-(e\rho /\theta_+)$, $\langle b^- \rangle =0$ and $\langle
A_0^{\rm in} \rangle =0$.

An expansion of the fermion determinants about the above mentioned
saddle point yields the on-loop effective action for the gauge
fields to be
\begin{eqnarray}
S_{\rm eff} &=&
   -{1 \over 2}\int d^3x \int d^3 x^\prime (A_\mu^\uparrow
   +a_\mu^+ +a_\mu^- +A_0^{\rm in}\delta_{\mu 0})(x)\Pi^{\mu\nu}_
   \uparrow (x,x^\prime)(A_\nu^\uparrow
 +a_\nu^+ +a_\nu^- +A_0^{\rm in}\delta_{\nu 0})(x^\prime) \nonumber \\
 & &   -{1 \over 2}\int d^3x \int d^3 x^\prime (A_\mu^\downarrow
   +a_\mu^+ -a_\mu^- +A_0^{\rm in}\delta_{\mu 0})(x)\Pi^{\mu\nu}_
   \downarrow (x,x^\prime)(A_\nu^\downarrow
 +a_\nu^+ -a_\nu^- +A_0^{\rm in}\delta_{\nu 0})(x^\prime) \nonumber \\
 & & +\int d^3x \left[ {\theta_+ \over 2 }\epsilon^{\mu\nu \lambda}
  a_\mu^+\partial_\nu a_\lambda^+ +
  {\theta_- \over 2 }\epsilon^{\mu\nu \lambda}
  a_\mu^-\partial_\nu a_\lambda^- \right]
    +{1 \over 2}\int d^3x \int d^3 x^\prime A_0^{\rm
   in}(x)V^{-1}(x-x^\prime)A_0^{\rm in} (x^\prime)\; .
  \nonumber \\
  & &
 \label{eq21}
\end{eqnarray}
The polarization tensors $\Pi^{\mu\nu}_{\uparrow , \downarrow}$
have the same form (in momentum space) as Eq.~(\ref{eq11}) with
the form factors in the lowest order in ${\bf q}^2$ to be
\begin{mathletters}
\label{eq22}
\begin{equation}
\Pi_0^{ \uparrow , \downarrow}= -\left( {e^2 \over 2\pi}\right)
{\bar{\omega}_c
\over \omega^2 -\bar{\omega}_c^2}
p_{ \uparrow , \downarrow } \; \; ; \; \;
\Pi_1^{ \uparrow , \downarrow}= \Pi_o^{\uparrow , \downarrow}
\bar{\omega}_c \; ;
\end{equation}
\begin{equation}
\Pi_2^\uparrow =  -{e^2 \over 4\pi m^\ast} \bar{\omega}_c^2
\left[ {3 \over \omega^2 -\bar{\omega}_c^2} -
{4 \over \omega^2 -4\bar{\omega}_c^2 } \right]
p_ \uparrow (p_ \uparrow -1)\; ;
\end{equation}
\begin{equation}
\Pi_2^\downarrow =  -{e^2 \over 4\pi m^\ast} \bar{\omega}_c^2
\left[ {3 \over \omega^2 -\bar{\omega}_c^2} -
{4 \over \omega^2 -4\bar{\omega}_c^2 } \right]
p_ \downarrow (p_ \downarrow +1) \; .
\end{equation}
\end{mathletters}
Note that $\Pi_0^\uparrow \neq \Pi_0^\downarrow $ any more.

The effective action for external probes $A_\mu^{\uparrow ,
\downarrow}$, which is obtained by the integration over internal
fields $a_\mu^+$, $a_\mu^-$ and $A_0^{\rm in}$ has  the same
form as Eq.~(\ref{eq13}). Now the density-density correlations
among different spin species $K^{00}_{rr^\prime}$ are obtained
(for small ${\bf q}^2$) as below:
\begin{mathletters}
\label{eq23}
\begin{eqnarray}
K^{00}_{\uparrow  \uparrow} &=& {{\bf q}^2 \over \Pi_0^\uparrow
+\Pi_0^\downarrow } \left[ \Pi_0^\uparrow \Pi_0^\downarrow
-{\Pi_0^{\uparrow 2} \theta_+^2 \over \left( \Pi_0^\uparrow +
\Pi_0^\downarrow \right) ^2 \omega^2 - \left( \Pi_1^\uparrow +
\Pi_1^\downarrow +\theta_+ \right) ^2 } \right]
+ {\cal O}(({\bf q}^2)^2) \; ; \\
K^{00}_{\downarrow  \downarrow} &=& {{\bf q}^2 \over \Pi_0^\uparrow
+\Pi_0^\downarrow } \left[ \Pi_0^\uparrow \Pi_0^\downarrow
-{\Pi_0^{\downarrow 2} \theta_+^2 \over \left( \Pi_0^\uparrow +
\Pi_0^\downarrow \right) ^2 \omega^2 - \left( \Pi_1^\uparrow +
\Pi_1^\downarrow +\theta_+ \right) ^2 } \right]
+ {\cal O}(({\bf q}^2)^2) \; ; \\
K^{00}_{\uparrow  \downarrow}= K^{00}_{\downarrow  \uparrow}
& =& -{{\bf q}^2 \over \Pi_0^\uparrow
+\Pi_0^\downarrow } \left[ \Pi_0^\uparrow \Pi_0^\downarrow
+{\Pi_0^\uparrow \Pi_0^\downarrow
\theta_+^2 \over \left( \Pi_0^\uparrow +
\Pi_0^\downarrow \right) ^2 \omega^2 - \left( \Pi_1^\uparrow +
\Pi_1^\downarrow +\theta_+ \right) ^2 } \right]
+ {\cal O}(({\bf q}^2)^2) \; ,
\end{eqnarray}
\end{mathletters}
in the limit $\theta_- \rightarrow  \infty \, (s_- =0)$ and
$\lim_{{\bf q}^2 \rightarrow  0} V({\bf q}^2) {\bf q}^2 =0$.

The generating functional, for equal-time density-density
correlations among mixed spins, is given by Eq.~(\ref{eq15})
with the following ${\cal K}$ matrix:
\begin{equation}
{\cal K}={{\bf q}^2 \over 2\pi}\left( {e^2 \over 2} \right) {1 \over
(p_ \uparrow + p_ \downarrow )} \left[
\begin{array}{cc}
p_ \uparrow p_ \downarrow + {p_ \uparrow ^2 \over 2s(p_ \uparrow
+p_ \downarrow )+1 } &
-p_ \uparrow p_ \downarrow + {p_ \uparrow  p_ \downarrow
\over 2s(p_ \uparrow
+p_ \downarrow )+1 } \\
-p_ \uparrow p_ \downarrow + {p_ \uparrow  p_ \downarrow
\over 2s(p_ \uparrow
+p_ \downarrow )+1 } &
p_ \uparrow p_ \downarrow + {p_ \downarrow ^2 \over 2s(p_ \uparrow
+p_ \downarrow )+1 }
\end{array}
\right]
\label{eq24}
\end{equation}

\section{Wave Functions}

Using the same procedure as employed by Lopez and Fradkin
\cite{lopez1}, we write the square of the modulus of the ground
state (non-degenerate) wave function for QH states of
mixed spin as
\begin{equation}
\left\vert \Psi [\rho_ \uparrow ,\rho_ \downarrow ]
\right\vert^2 = \int [dA_0^\uparrow ][dA_0^\downarrow ] {\cal Z}
[A_0^\uparrow , A_0^\downarrow ] \exp \left[ -ie \int {d^2 {\bf
q} \over (2\pi)^2} A_0^r ({\bf q})\delta\rho_r (-{\bf q})
\right]
 \; , \label{eq25}
\end{equation}
where $\delta\rho_r ({\bf q})$ is the fourier transform of the density
fluctuation
\begin{equation}
\delta\rho_r (X) =\sum_{i=1}^{N_r}\delta (X-X_i^r) -\rho_r \; ,
\label{eq26}
\end{equation}
where $N_r$ is the number of electrons with spin index $r (=
\uparrow , \downarrow )$ and $\rho_r$ is the
corresponding mean density. Now the integrations over
$A_0^\uparrow $ and $A_0^\downarrow $ in Eq.~(\ref{eq25}) yield
\begin{equation}
\left\vert \Psi [\rho_ \uparrow ,\rho_ \downarrow ]
\right\vert^2 = \exp \left[ {e^2 \over 2} \int {d^2 {\bf q}
\over
(2\pi)^2} \delta\rho_r ({\bf q}) {\cal K}_{rr^\prime}^{-1}
\delta\rho_{r^\prime} (-{\bf q}) \right] \; ,
\label{eq27}
\end{equation}
where ${\cal K}_{rr^\prime}$ are given by Eqs.~(\ref{eq17} and
\ref{eq24}) for two different cases.

\subsection{Unpolarized States}

For the unpolarized states, both up and down spins are equally
populated. Therefore, $\rho_ \uparrow  = \rho_ \downarrow = \rho
/2 $ and $N_ \uparrow = N_ \downarrow =N/2$ where $N$ is the
total number of particles. Transforming Eq.~(\ref{eq27}) back
into the real space and using Eq.~(\ref{eq17}) for this case, we
find
\begin{eqnarray}
\left\vert \Psi [\rho_ \uparrow ,\rho_ \downarrow ]
\right\vert^2 &=&
\exp \left[ {2sp +1 \over p} \int d^2X \int
d^2Y \delta\rho_ \uparrow (X) \ln { \left\vert X-Y \right\vert
\over R} \delta\rho_ \uparrow (Y) \right] \nonumber \\
& & \times \exp \left[ {2sp +1 \over p} \int d^2X \int
d^2Y \delta\rho_ \downarrow (X) \ln { \left\vert X-Y \right\vert
\over R} \delta\rho_ \downarrow (Y) \right] \nonumber  \\
& & \times \exp \left[ 2(2s) \int d^2X \int
d^2Y \delta\rho_ \uparrow (X) \ln { \left\vert X-Y \right\vert
\over R} \delta\rho_ \downarrow (Y) \right]
\label{eq28}
\end{eqnarray}
where $R$ is the radius of the system which serves as a
long-distance cut-off such that the magnetic length and
inter-electron distance are much less than $R$. Thus, from
Eqs.~(\ref{eq26} and \ref{eq28}), we obtain the modulus square
of the wave function for the ground state of filling fraction
(\ref{eq8}),
\begin{eqnarray}
\left\vert \Psi \left( X_1^\uparrow ,\cdots ,X_{N/2}^\uparrow ,
 X_1^\downarrow , \cdots ,X_{N/2}^\downarrow \right)
\right\vert^2  &=& \prod_{i<j}^{N/2} \left\vert X_i^\uparrow -
X_j^\uparrow \right\vert^{2(2sp+1)/p} \nonumber \\
& & \times \prod_{k<l}^{N/2} \left\vert X_k^\downarrow -
X_l^\downarrow \right\vert^{2(2sp+1)/p} \nonumber
\prod_{i,k}^{N/2} \left\vert X_i^\uparrow -
X_k^\downarrow \right\vert^{2(2s)}  \nonumber \\
& & \times \exp \left[ -{1 \over 2l_0^2}
\left( \sum_{i=1}^{N/2} \left\vert
X_i^\uparrow \right\vert^2 +\sum_{k=1}^{N/2} \left\vert
X_k^\downarrow \right\vert^2 \right) \right] \; ,
\label{eq29}
\end{eqnarray}
where $X_i^\uparrow (X_i^\downarrow )$ represents the co-ordinate
of i-th  spin-up (down) particle and $l_0 = (eB)^{-1/2}$ is the
magnetic length.

Eq.~(\ref{eq29}) reproduces the wave functions for unpolarized
QH states at filling fractions
(i) $\nu =2 (s=0,p=1)$ and, (ii) $\nu ={2 \over 5}
(s=1,p=1)$ as proposed by Halperin \cite{halp}. It agrees with
the trial wave functions proposed by Jain et al. \cite{jain2,wu}.
However, the wave functions differ (except the state $\nu = {2
\over 5}$) from the trial wave functions for the sequence $\nu =
2n/(3n+2)$, proposed by Belkhir and Jain \cite{belk}.
Note that while the infra red cut-off is required to capture the
exponential part,
the Jastrow part of the wave function is
obtained unambigiously. Clearly, the wave functions are analytic
for all states with $\nu = 2p/(4sp+1)$ for $p=\pm 1$. On the other
hand, all the states with finite $p$ are non-analytic.
However, the even denominator states $(p = \infty)$ are again analytic
in nature. The reason behind analyticity/non-analyticity will be
discussed below. Interestingly, for all the integer states $(s=0)$, the
exponent of the Jastrow form $\vert X_i^\uparrow -
X_k^\downarrow \vert $ vanishes.
Therefore the model predicts that the particles with unlike
spins are uncorrelated for the integer states.

\subsection{Partially/Fully Polarized States}

Partially polarized states have unequal population in spin
species. As discussed earlier, for small Zeeman energy, let $p_
\uparrow = p_ \downarrow +1 =p$ (say). Hence, $\rho_ \uparrow
=(p/(2p-1))\rho$, $\rho_ \downarrow = ((p-1)/(2p-1))\rho$, $N_
\uparrow =(p/(2p-1))N$ and $N_ \downarrow = ((p-1)/(2p-1))N$.
Using the same procedure as discussed for unpolarized
states in section (III-A), we obtain the wave function for
partially/fully polarized states from Eqs.~({\ref{eq24},
\ref{eq26} and \ref{eq27}),
\begin{eqnarray}
\left\vert \Psi \left( X_1^\uparrow ,\cdots ,
X_{N_ \uparrow }^\uparrow ,
X_1^\downarrow , \cdots , X_{N_ \downarrow }^\downarrow \right)
\right\vert^2  &=& \prod_{i<j}^{N_ \uparrow }
\left\vert X_i^\uparrow -
X_j^\uparrow \right\vert^{2(2sp+1)/p} \nonumber \\
& & \times \prod_{k<l}^{N_ \downarrow } \left\vert X_k^\downarrow -
X_l^\downarrow \right\vert^{2(1+2s(p-1))/(p-1)}
\prod_{i,k}^{N_ \uparrow , N_ \downarrow }
\left\vert X_i^\uparrow -
X_k^\downarrow \right\vert^{2(2s)}  \nonumber \\
 & & \times \exp \left[ -{1 \over 2l_0^2} \left(
\sum_{i=1}^{N_ \uparrow } \left\vert
X_i^\uparrow \right\vert^2 +\sum_{k=1}^{N_ \downarrow } \left\vert
X_k^\downarrow \right\vert^2 \right) \right] \; .
\label{eq30}
\end{eqnarray}

It is clear that the wave functions are non-analytic for
$p>1$, which is the case for partial polarization.
However, for $p=1$ which essentially gives fully polarized
Laughlin sequence \cite{laugh}, (\ref{eq30}) is
in agreement with the result of
Lopez and Fradkin \cite{lopez1} for spinless system.

\subsection{Source of Analyticity/non-analyticity}

Let us write the Jastrow part of the wave function for unpolarized
states (\ref{eq29}) in the form,
\begin{eqnarray}
\left\vert \Psi \right\vert^2 &\approx & \prod_{i<j}^{N/2}
\left\vert X_i^\uparrow -X_j^\uparrow \right\vert^{2\left( (1/
\theta)+ (1/\Pi_1^\uparrow ) \right) e^2/(2\pi)}
\prod_{k<l}^{N/2}
\left\vert X_k^\downarrow -X_l^\downarrow \right\vert^{2\left( (1/
\theta)+ (1/\Pi_1^\downarrow ) \right) e^2/(2\pi)} \nonumber \\
& & \times \prod_{i,k}^{N/2}
\left\vert X_i^\uparrow -X_k^\downarrow \right\vert^{2\left( 1/
\theta \right) e^2/(2\pi)}       \; ,
\label{eq31}
\end{eqnarray}
where $\Pi_1^{\uparrow , \downarrow}$ are evaluated at $\omega =0\, ,\,
{\bf q}^2 =0$. Note that the exponents
of the Jastrow forms are determined
by the parity and time reversal violating factors of the
effective action given by Eqs.~(\ref{eq10}, \ref{eq11}) .
$ 1/\theta$ is always an even integral multiple of $(2\pi/e^2)$, and
depending on the effective filling factor $(p)$,
$1/\Pi_1^{\uparrow , \downarrow}$ should either be an integral or
fractional multiple of $(2\pi /e^2)$. Therefore the wave
function becomes analytic when $1/\Pi_1^{\uparrow , \downarrow}$
is an integral multiple of $(2\pi /e^2)$, i.e., only for $p =\pm
1$, while for other values of $p$, $1/\Pi_1^{\uparrow ,
\downarrow}$ is fractional multiple of $(2\pi /e^2)$ and hence
the wave function becomes non-analytic.

Indeed, the exponents describe the number of effective vortices
associated with a particle which is seen by others and therefore the
exponents of the Jastrow form between like spin particles differ
from the same between unlike spins.
It is natural that $\Psi $ should reflect the nature of vortices
associated with the fermion.
$\Psi$ is determined from the density-density correlations which
represent, in fact,
the change in local density of the system and hence
the change in CS magnetic field. This causes a
change in the local current
which is represented by the vortices.

A similar argument for analyticity/non-analyticity also holds
for partially/fully polarized QH states.

\section{Kohn's mode and Hall Conductivity}

We have introduced the external probes $A_\mu^{\uparrow ,
\downarrow}$, mainly for the computational purposes, {\it viz.},
for determining the mixed spin
density-density correlations which are used to
determine the ground state wave functions of QH states. All the
electro-magnetic responses of the system are determined by the
physical electro-magnetic probe $A_\mu$ which
couples to both the spins. It is not necessary to compute the
response function {\it de novo}, since, the correlations
found from the probes $A_\mu^{\uparrow , \downarrow}$ are
related to that from the probe $A_\mu$.
If we write
\begin{equation}
S_{\rm eff} [A_\mu ] = {1 \over 2}\int {d^3 q \over (2\pi)^3}
A_\mu (q) K^{\mu\nu} A_\nu (-q) \; ,
\label{eq32}
\end{equation}
the electromagnetic response tensor
$K^{\mu\nu}$ is related to $K^{\mu\nu}_{rr^\prime}$ (which
we have evaluated with the probes
$A_\mu^{\uparrow , \downarrow}$), through
\begin{equation}
K^{\mu\nu}=\sum_{r,r^\prime} K^{\mu\nu}_{rr^\prime} \; .
\label{eq33}
\end{equation}
Having thus determined $K^{\mu\nu}$,
considering translational and gauge invariance, we write
\begin{equation}
K^{\mu\nu}=K_0(q^2g^{\mu\nu}-q^\mu q^\nu)+(K_2 -K_0)({\bf q}^2
\delta^{ij}-q^iq^j)\delta^{\mu i}\delta^{\nu j} +iK_1
\epsilon^{\mu \nu \lambda}q_ \lambda  \; ,
\label{eq34}
\end{equation}
where $K_0$, $K_1$ and  $K_2$ are functions of $\omega$ and
${\bf q}^2$.

The density-density correlation function can then be evaluated
in the limit ${\bf q}^2 \rightarrow  0$ as
\begin{equation}
K^{00} (\omega \, ,\, {\bf q}^2) \equiv -K_0{\bf q}^2 = -({e^2
\rho \over m^\ast}){{\bf q}^2 \over \omega^2 -\omega_c^2} +{\cal
O}(({\bf q}^2)^2)
\label{eq35}
\end{equation}
which is the same
for both the cases, i.e., unpolarized and
polarized states. Note that the pole occurring in the
density-density correlation function is at $\omega = \omega_c$,
which is the cyclotron frequency due to the applied field only.
This is the well known as Kohn's mode \cite{kohn} which has got restored
by the fluctuation of CS gauge fields over their mean values.
Further, to the leading order in ${\bf q}^2$, $K^{00}$ also
saturates the {\it f}-sum rule as Lopez and Fradkin
\cite{lopez3} have observed for the spinless case.

We may now
obtain the Hall conductivity of the system to be
\begin{equation}
\sigma_H \equiv K_1 (0,0) =\left\{
\begin{array}{ll}
      {2 \Pi_1 (0,0)\theta \over 2\Pi_1 (0,0)+
\theta}\; , & \mbox{for unpolarized states} \\
  {\left(\Pi_1^\uparrow (0,0)+\Pi_1^\downarrow (0,0)
 \right)\theta_+ \over \left( \Pi_1^\uparrow
 (0,0)+\Pi_1^\downarrow (0,0) \right) +\theta_+ }\; , & \mbox{for
 partially polarized states}
 \end{array}
 \right. \; .
 \label{eq36}
 \end{equation}
 Therefore $\sigma_H = \nu (e^2 /2\pi)$, which is quantized for
 the corresponding filling fractions $\nu$ (\ref{eq8},
 \ref{eq20}).

\section{conclusion}

In conclusion, we have determined the wave functions for
arbitrarily polarized QH states within the doublet model
\cite{spin} proposed recently. Our findings reduce to that of
Lopez and Fradkin \cite{lopez1} for fully polarized states. We
are able to recover the wave function proposed by Halperin
\cite{halp} for $\nu = {2 \over 5}$. Our wave functions do not
agree with those of Belkhir and Jain \cite{belk}, except for
$\nu ={2 \over 5}$. This disagreement is not surprising since
the sequence for unpolarized states obtained here is different
from the sequence $\nu =2n/(3n+2)$ employed by Belkhir and Jain
\cite{belk}
to write their wave functions. We have also confirmed by an
explicit One-loop computation that the Hall conductivity does
indeed get quantized at those filling fractions that follow from
the model. Finally, we have given a physical picture for the
non-analytic nature of the wave functions, and shown that quantum
fluctuations restore the Kohn mode.
It would be interesting if spin correlations are
measured and compared with the findings here.
It would be equally
interesting if the other states corresponding to two gaps of
excitation are also discovered experimentally. One simple way of
identifying them would be by measuring the activation in
diagonal resistivity.
Finally, there remains the possibility of multilayered
arbitrarily polarized QH states.
A study of such systems would require a fusion of the model
used here with that of Lopez and Fradkin \cite{lopez4} which
discusses double layered spinless fermion (fully polarized)
systems. We believe
that this fusion might well be possible since the two models
are so very similar to each other.

\section*{Acknowledgment}
We thank the referee for bringing the paper of Lopez and
Fradkin \cite{lopez4}to our notice.

\appendix

\section*{Comparision with bilayered systems}

Recently Lopez and Fradkin (LF) \cite{lopez4} have studied QH
effects in bilayered systems with complete spin polarization.
There is a close resemblance between their approach with the
one we have taken here: The Lagrangians are formally
the same, with a fermion doublet (while there is layered index
in LF, we have the spin index here) and a matrix valued CS strength.
In bilayered system the interlayer and intralayer interaction
potentials are different, unlike in the case of
spin-1/2 fermions for which the interaction potential does not
depend on spin. However, this aspect is not important
as long as we consider short-range interactions.

The crucial difference, in fact, is in the (physical) choice of
$\Theta$. In their work, LF \cite{lopez4} have chosen the
$\Theta $ matrix as
\begin{equation}
\Theta_{{\rm LF}}= \frac{e^2}{2\pi}\frac{1}{4s_1 s_2 -n^2}\left(
\begin{array}{cc}
2s_2 & -n \\
-n & 2s_1 \end{array} \right) \; ,
\label{eqa1}
\end{equation}
where $s_1 , s_2 $ and $n$ are integers. Clearly $\Theta_{{\rm
LF}}$ has more parameters compared to our choice of $\Theta$ in
Eq.~(\ref{eq4}) since LF have assumed more generically that the
particles in two different layers feel unequal number of flux
quanta which is, in fact, not the case for spin-1/2 fermions in
a single layer.
In contrast, our case corresponds to $s_1 =s_2$. Note that this
choice is a strict requirement that the parity operation
transforms the up-spin to down-spin electron and vice versa.
There is no such requirement in the LF case. More importantly,
note that $\Theta_{{\rm LF}}$ is ill-defined for
$s_1 =s_2 =n/2 $.
As a consequence, there are fundamental differences in results
and interpretation in the two approaches which we discuss below.

Consider the spin unpolarized (singlet) state first. The corresponding
sequence of states obtained by LF \cite{lopez4} for equal
population in the two layers is identical to Eq.~(\ref{eq8}) here
for the choice of $s_1 =s_2 =n/2$ in Eq.~(\ref{eqa1}).
A closer look however shows that it is precisely for these states,
(characterized by the filling fractions $\nu_1 =\nu_2$ in  the two
layers and the number of particles $N_1=N_2$ in two layers in
Ref.~\cite{lopez4}) that the CS strength $\Theta_{{\rm LF}}$
becomes ill-defined. The ensuing dynamics is also ill-defined.
Indeed as we have quoted earlier,
LF \cite{lopez4} point out in their paper ``the spin
singlet state (3,3,2), which has filling fraction $\nu =2/5$,
cannot be described within the Abelian Chern-Simons approach".
In fact, no spin unpolarized states given by filling fraction in
(\ref{eq8}), can be described in the approach of LF
\cite{lopez4}.
In contrast, our $\Theta$ is given by
\begin{equation}
\Theta =\frac{e^2}{2\pi}\frac{1}{4s} \left(
\begin{array}{cc}
1 & 1 \\
1 & 1 \end{array} \right) \; ,
\label{eqa2}
\end{equation}
corresponding to $\theta_1 =\theta_2 $, with eigen values
$\theta_+ =(e^2 /2\pi )(1/2s)$ and $\theta_- =0$, which keeps
the composite fermion picture intact.
This well defined matrix naturally leads to
unpolarized states (See Ref.~\cite{spin}).
The dynamics is also well defined, allowing
us to obtain quantization of Hall conductivity (\ref{eq36}) and
many-body wave functions (\ref{eq29}) for these states.
Again, as we have stated earlier, we recover Halperin wave
function \cite{halp} for spin unpolarized $\nu =2/5$ state.
Note that since all the elements in $\Theta $ in Eq.~(\ref{eqa2})
are same, the flux seen by all the particles are same
irrespective of their spins, unlike $\Theta_{{\rm LF}}$ for
which flux seen by paricles in the same layer are different from
particles in different layers.

In our approach, $\nu =1/m$ ($m$ odd) states (which are the
Laughlin sequence) are always fully polarized (see Eq.~
\ref{eq20}). This, in fact, is true as it is seen both in
the experiment \cite{engel} and numerical calculations
\cite{tchak1} in a single layer.
Contrarily in the work of LF \cite{lopez4}, these states can also
be obtained for equal population of particles in the two
layers corresponding to filling fractions $\nu_1 =\nu_2 =1/2m $
(where 1 and 2 refer to two different layers). The wave functions
for fully polarized $\nu =1/m $ states are well known Laughlin
wave functions which is also derived here (See Eq.~(\ref{eq30})
for fully polarized limit). On the other hand, LF have obtained
the wave functions for $\nu =1/m$ states, in their above
construction, as $(m,m,m)$ Halperin wave functions with an
extra non-analytic piece (which we do not encounter)
due to the presence of a gapless mode in the
spectrum of collective excitations for these states.

The $\nu =1/2$ state in bilayered systems corresponds to the
assignment of filling fractions $\nu_1 =\nu_2 =1/4$ in two
layers which yield the gaps $\bar{\omega}_c^{1,2} =\omega_c /4$.
With their choice, LF obtain the wave function for which coincides
with the one obtained by numerical computation in double layers. On the
other hand, the present model yields
$\bar{\omega}_c^{\uparrow ,
\downarrow} =0$. This is again closer to several  experiments
\cite{will,kang,gold} which have unambigiously
verified that $\bar{\omega}_c =0$
for $\nu =1/2$ in a single layer.

It is of course not that the two models are entirely different
in all aspects.
Indeed, the double layered systems
with dissimilar gaps in two layers
are exact analogues
of the spin systems with dissimilar gaps for spin up and down
states because $\Theta_{{\rm LF}}$ becomes identical to $\Theta
$ here, provided one puts $s_1 =s_2$ in Eq.~(\ref{eqa1}).
Note, however, that spin unpolarized states (which are seen
experimentally \cite{clark,eisen1,eisen2,engel}) can never arise
if $\bar{\omega}_c^\uparrow \neq \bar{\omega}_c^\downarrow $.
Experimentally observed partially polarized or fully polarized
states which are given by Eq.~(\ref{eq19})
also correspond to $\bar{\omega}_c^\uparrow
 = \bar{\omega}_c^\downarrow $.
It will be interesting to observe experimentally whether there
is any such states for which
$\bar{\omega}_c^\uparrow \neq \bar{\omega}_c^\downarrow $
in which case there would be a one-to-one correspondence between
the two models.

Finally, we remark that LF \cite{lopez4} have also studied spin
unpolarized states in a single layer by employing a non-abelian
CS interaction. In their picture, electrons are composite of
holons and spinons. Charged spinless holons interact with U(1)
CS gauge field where as neutral spin-1/2 spinons interact with
SU(2) CS gauge field. They both obey semionic statistics --
which indicates a departure from composite fermion model
(where spin and charge are not seperated) to which we completely
adhere.


\begin{references}

\bibitem[*]{e1} e-mail: ssman@iitk.ernet.in
\bibitem[**]{e2} e-mail: vravi@iitk.ernet.in
\bibitem{frad} E. Fradkin, Nucl. Phys. {\bf B 389}, 587 (1993).
\bibitem{lopez1} A. Lopez and E. Fradkin, Phys. Rev. Lett. {\bf 69},
		2126 (1992).
\bibitem{jain1} J. K. Jain, Phys. Rev. Lett. {\bf 63}, 199 (1989).
\bibitem{jain2} J. K. Jain, Phys. Rev. {\bf B 41}, 7653 (1990).
\bibitem{laugh} R. B. Laughlin, Phys. Rev. Lett. {\bf 50}, 1395 (1983).
\bibitem{trug} S. Trugman and S. Kivelson, Phys. Rev. {\bf B 31},
	       5280 (1985).
\bibitem{hald} F. D. M. Haldane, in {\it The Quantum Hall Effect},
	       edited by R. Prange and S. Girvin (Springer-Verlag,
	       1990).
\bibitem{halp} B. I. Halperin, Helv. Phys. Acta {\bf 56}, 75 (1983).
\bibitem{clark} R. G. Clark, S. R. Haynes, A. M. Suckling,
		J. R. Mallett, P. W. Wright, J. J. Harris
	      and C. T. Foxon, Phys. Rev. Lett. {\bf 62}, 1536 (1989).
\bibitem{eisen1} J. P. Eisenstein, H. L. Stormer, L. N. Pfeiffer
		and K. W. West, Phys. Rev. Lett. {\bf 62}, 1540 (1989).
\bibitem{eisen2} J. P. Eisenstein, H. L. Stormer, L. N. Pfeiffer
		and K. W. West,
		Phys. Rev. {\bf B 41}, 7910 (1990).
\bibitem{engel} L. W. Engel, S. W. Hwang, T. Sajoto, D. C. Tsui and
               M. Shayegan, Phys. Rev. {\bf B 45}, 3418 (1992).
\bibitem{tchak1} T. Chakraborty and F. C. Zhang, Phys. Rev.
	   {\bf B 29}, 7032 (1984); F. C. Zhang and T. Chakraborty,
		   {\it ibid.} {\bf 30}, 7320 (1984); E. H. Rezayi,
		   {\it ibid.} {\bf 36}, 5454 (1987).
\bibitem{tchak2} P. A. Maksym, J. Phys. Condens. Matter {\bf 1},
		     6229 (1989); T. Chakraborty and P. Pietilainen,
		     Phys. Rev. {\bf B 41}, 10862 (1990).
\bibitem{wu} X. G. Wu, G. Dev and J. K. Jain, Phys. Rev. Lett.
		{\bf 71}, 153 (1993).
\bibitem{belk} L. Belkhir and J. K. Jain, Phys. Rev. Lett. {\bf 70},
		   643 (1992).
\bibitem{spin} S. S. Mandal and V. Ravishankar, preprint cond-mat
		   9506071; submitted to Phys. Rev. Lett.
\bibitem{wen} Models with matrix valued coupling have been
     considered earlier, in a different context by X. G. Wen and A.
        Zee, Phys. Rev. {\bf B 46}, 2290 (1992).
\bibitem{lopez2} A. Lopez and E. Fradkin, Phys. Rev. {\bf B 44},
		    5246 (1991).
\bibitem{fnote1} These are introduced merely for the calculation
		    purposes, viz, the mixed spin density-density
		    correlations. However, any electromagnetic
		    response of the system has to be determined with
		    the real electro-magnetic probe, as we do in
		    section IV.
\bibitem{kohn} W. Kohn, Phys. Rev. {\bf 123}, 1242 (1961).
\bibitem{lopez3} A. Lopez and E. Fradkin, Phys. Rev. {\bf B 47},
		    7080 (1993).
\bibitem{lopez4} A. Lopez and E. Fradkin, Phys. Rev. {\bf B 51},
		    4347 (1995).
\bibitem{frohl} J. Frohlich, T. Kerler and P. A. Marchetti,
		   Nucl. Phys. {\bf B 374}, 111 (1992).
\bibitem{will} R. L. Willett, R. R. Ruel, K. W. West and L. N.
                  Pfeiffer, Phys. Rev. Lett. {\bf 71},
		   3846 (1993).
 \bibitem{kang} W. Kang, H. L. Stormer, L. N. Pfeiffer,
                K. W. Baldwin and K. W. West, Phys. Rev. Lett. {\bf 71},
			3850 (1993).
 \bibitem{gold} V. J. Goldman, B. Su and J. K. Jain, Phys. Rev. Lett.
			{\bf 72}, 2065 (1994).

\end{references}
\end{document}